\documentclass[sigchi-a]{acmart}

\usepackage{todonotes, soul}
\def\BibTeX{{\rm B\kern-.05em{\sc i\kern-.025em b}\kern-.08emT\kern-.1667em\lower.7ex\hbox{E}\kern-.125emX}}

\begin{document}

%
\title{Multi-Modal Measurements of Mental Load}

%
\author{Ingo Keller}
\email{ijk1@hw.ac.uk}
\author{Muneeb Imtiaz Ahmad}
\email{m.ahmad@hw.ac.uk}
\author{Katrin Lohan}
\email{k.lohan@hw.ac.uk}
\affiliation{%
  \institution{ Social Robotics Group, School for Mathematical and Computer Sciences, Heriot-Watt University}
  \streetaddress{P.O. Box 1212}
  \city{Edinburgh}
  \state{United Kingdom}
}

%
\renewcommand{\shortauthors}{Keller et al.}

%
\begin{abstract}

This position paper describes an experiment conducted to understand the relationships between different physiological measures including pupil Diameter, Blinking Rate, Heart Rate, and Heart Rate Variability in order to develop an estimation of users' mental load in real-time (see Sidebar \ref{fig:experiment}). Our experiment involved performing a task to spot a correct or an incorrect word or sentence with different difficulties in order to induce mental load. We briefly present the analysis of task performance and response time for the items of the experiment task.


\end{abstract}

%
%


%

\begin{sidebar}
    \includegraphics[width=.5\textwidth]{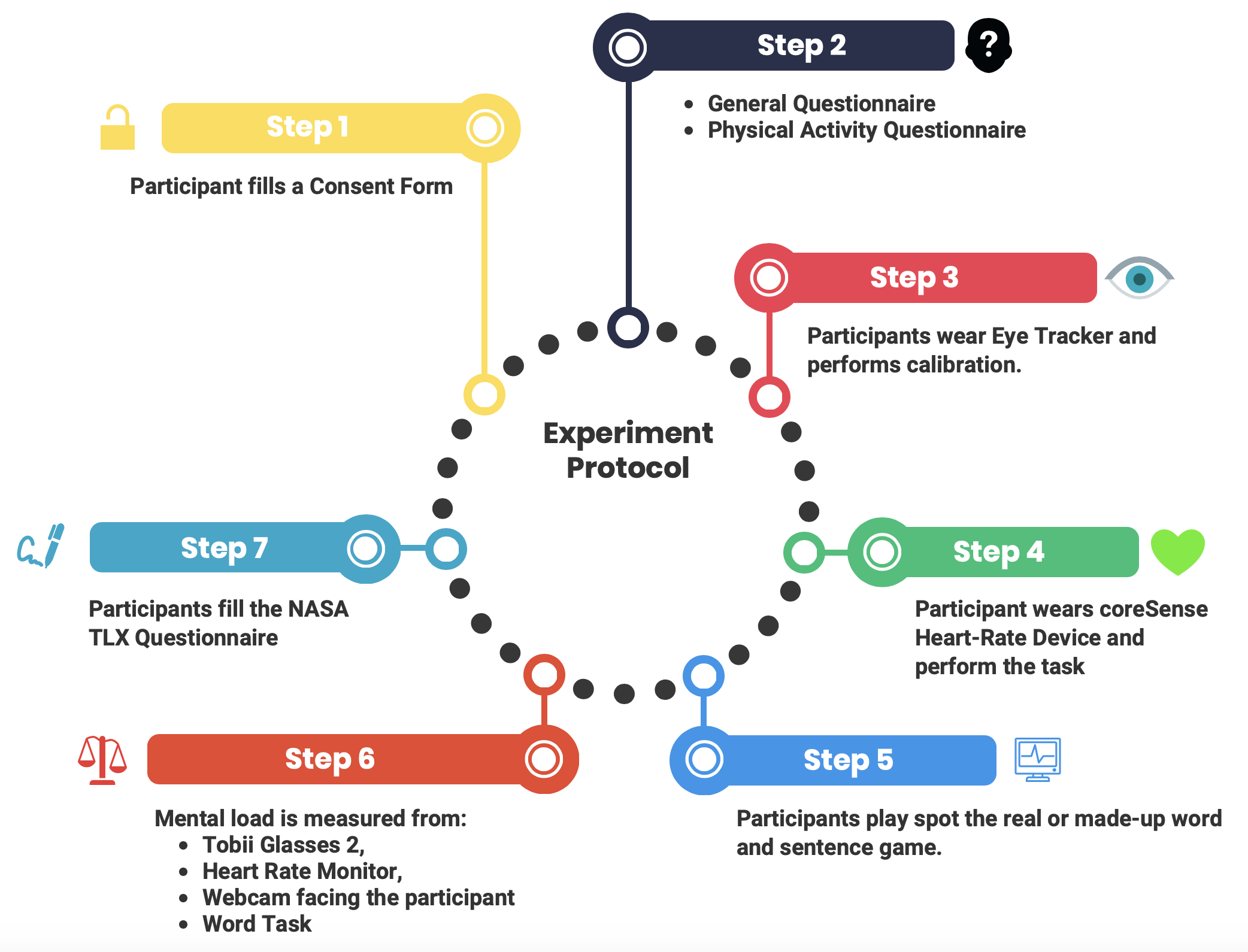}
    \caption{Experimental Protocol}
    \label{fig:experiment}
\end{sidebar}

%

%
\maketitle
\section{Introduction}

There is an ongoing interest in creating robots that can adapt their behaviour according to users traits in the field of Human-Robot/Computer Interaction (HCI/HRI) \cite{ahmad2017systematic}. User's Cognitive Load or Operator's Mental Load are among important attributes during the HCI. It has been found that a higher level of cognitive or mental load of the users may have an adverse effect on their task performance during HCI scenario \cite{reeves2004guidelines}.

The amount of effort placed on a user's working memory during a task is referred to as the cognitive load.  According to the Cognitive Load Theory \cite{sweller1994cognitive}, there are three different types of load produced during problem-solving or learning: 1) intrinsic load, 2) the extraneous load and 3) the germane load.
\begin{enumerate}
    \item Intrinsic load depends on the complexity of the structure of the material and its association with the learner.
    \item The extraneous load is caused by the way this material is presented to the learner.
    \item The germane load is a result of the learner's ability to assimilate the material.
\end{enumerate}

On the other hand, cognitive load is also defined as a construct that can be measured in three dimensions: 1) mental load, 2) mental effort and 3) performance \cite{sweller1998cognitive}. In other words, past research indicates that cognitive load and mental load are related to each other in terms of the working memory. 
Therefore, we will be using them interchangeably. It is also interesting to note that the amount of workload varies between individuals \cite{baddeley1992working}. This suggests that the amount of working memory is different for each individual. Thus, it highlights the significance of measuring the amount of the mental load in real-time in order to adapt the Interface during the HRI or HCI. Consequently, we are also involved in creating the technology that can adapt to the user's mental load 
in real-time. 
More specifically, we want to personalise the behaviour of the robot or adjust the information on the user interface 
to adjust humans mental overload. 

It is currently an open challenge to measure the individual's mental load in a non-intrusive and robust manner in real-time \cite{paas2003cognitive,chen2012multimodal}. 
Existing literature shows that cognitive load can be measured in three different ways: through subjective rating questionnaires (NASA Task Load Index\footnote{ NASA Task Load Index - https://humansystems.arc.nasa.gov/groups/TLX/
downloads/TLXScale.pdf}), through physiological sensors, or through performance-based objective measures (Mathematical Equations). However, subjective rating and performance-based objective measures are limited and cannot be utilised in real-time during HCI or HRI to adjust the interface or the behaviour of the robot. 
On the contrary, existing literature has highlighted different physiological measures that are continuous and can be used to estimate mental load in real-time \cite{paas2003cognitive}. These measures include Pupil Diameter (PD), Blinking Rate (BR), Heart Rate (HR), Heart Rate Variability (HRV),  Electroencephalography (EEG) and Galvanic Skin Response (GSR). Each of these methods has been shown to measure mental load during various tasks; however, each of them has its limitations. For instance: Eye activity (PD, BR) may not be suitable for tasks requiring continuous reading. Similarly, the sensitivity of pupil changes in cognitive load diminishes with age \cite{van2004memory}. Additionally, the HRV may depend on the physical fitness of the individuals \cite{luque2013cognitive}. It is, therefore, significant to understand the relation between these physiological behaviours under different tasks or circumstances 
with each other and also onto each other. As our long-term goal is to estimate the user's mental load in real-time to personalise interfaces or robot's behaviour, therefore, we believe that understanding the relationship between different physiological behaviour is vital and will help us in achieving our long-term goal of creating a model that takes various physiological behaviours as input to estimate mental overload or cognitive load in an efficient manner.

To achieve our long-term goal, we are currently collecting data from various physiological sensors to understand the relationship between aforementioned physiological behaviours under different task during HCI and HRI respectively \cite{a,b,c}. Building on the prior work, 
this position paper describes another series of one of our experiments conducted to understand the relationship between different physiological behaviours during a game-based task. We intend to collect data based on various physiological behaviours (HR, HRV, BR, and PD) during different tasks and later want to create datasets based on these behaviours. Our long-term goal is to use these datasets for a linear mixed-effects regression model to estimate mental or cognitive load in real-time. 


\section{Related Work}

In the past, empirical findings have highlighted a relationship between mental load or cognitive load and various physiological behaviour (PD, HRV, HR, BR, GSR and EEG).  It has been shown that the HRV reduces with an increase in the amount of cognitive load during a range of computerised tasks \cite{mukherjee2011sensitivity}. Similarly, another study observed an increase in HR with an increase in the difficulty level of the task or in other words in a situation demanding higher cognitive load \cite{cranford2014measuring}. In addition, we find a body of literature that suggests that pupil diameter increases with an increase in the number of mental load \cite{reilly2018human, AI-HRI}. Moreover, research indicates that the average number of blinks per minute in normal healthy adults varies according to the task. A study conducted with 150 adults showed that individuals on average blink four to five times per minute during a reading activity. Similarly, the rate of blinking per minute is higher during resting (17) and during conversation (26) \cite{bentivoglio1997analysis}. Later, it has been shown that the blinking rate is minimised in the case of higher mental overload \cite{holland1972blinking}. 
\\
In summary, we find enough empirical evidence from past research implying that the change in the physiological behaviour can be attributed to cognitive load or mental load and it also refers to various levels of mental processing. It is also important to note that the data collected on physiological behaviours through the various state of the art sensors is not only continuous but is also a robust and accurate representation of the particular behaviour \cite{EliteHRV,TobiiProGlasses2}. 
It has been highlighted that users are required to wear cumbersome equipment; however, with the advancement of design and technology, existing solutions have been devised to collect such data in non-invasive ways. 

In relation to our long-term goal, we currently find empirical evidence that few of the physiological behaviours are correlated with each other. For instance, both pupil dilation and blink-eye rate were seen to be correlating with each other in a digit-sorting task \cite{siegle2008blink}. However, to the best of our knowledge, the relationship between all these behaviours have not been explored and is necessary to efficiently create an estimation for cognitive load or mental load. As highlighted we need to understand the relationships between different behaviours because this can help in creating a linear mixed-effect regression model that may estimate mental load in real-time during HCI or HRI. Consequently, We, in this paper, presents intial finding of a study conducted in order to collect data on these behaviours in a synchronous way to help us analyze the relationships between different physiological behaviours and later create a dataset that can be used to estimate cognitive load in real-time. 

\vspace{1em}
\section{Study}
\vspace{.5em}

The purpose of the study was to understand the relationship between different physiological measures (HR, HRV, PD and BR) to use these variables to help estimate mental load in real-time. 

\vspace{1em}
\subsection{Experiment Protocol}
\vspace{.5em}

The experiment was conducted with 41 participants (demographics as shown in table 1) in seven different steps:

\begin{itemize}
    \item Participant fills a Consent Form.
    \item Participant fills a General Questionnaire to report information on age, numbers of languages they speak and in case they have reading difficulties and a Physical Activity Questionnaire \cite{fogelholm2006international} to control for any bias in the HR and HRV measurements. 
    \item Participant wears Tobii Eye Tracker and performs a simple Calibration to get the values for their Pupil Diameter.
    \item Participant wears corSense  Heart-Rate Device and performs the task.
    \item Participant plays spot the real or made-up word and sentence game.
    \item Mental load is measured from Tobii Glasses 2 (BR, PD), Heart Rate Monitor (HR, HRV), and Webcam facing the participant during the game task.
    \item Participant fills the NASA TLX Questionnaire. 
\end{itemize}

\begin{sidebar}
    \includegraphics[width=.5\textwidth]{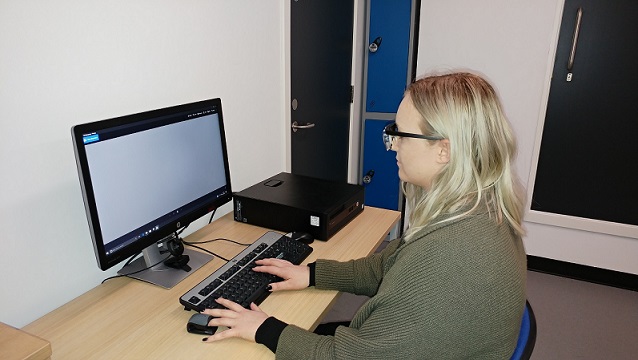}
    \caption{Setup: Participant is performing the game based task}
    \label{fig:cl_load}
\end{sidebar}

\vspace{1em}
\subsection{Task}

The task given was a correct or an incorrect English word or sentence recognition task with different difficulties. The participants were presented with items from the following categories. All word items were 10 letters long. All sentence items consisted of 10 words each. \\
\begin{center}
    \begin{tabular}{ | c | l | }
        \hline
        Item Type & Content                                                 \\
        \hline
        1 & a correct English word                                          \\
        2 & as 1 but with the middle letters switched                       \\
        3 & as 1 but with scrambled letters                                 \\
        4 & an arbitrary mnemonic word                                      \\
        5 & a correct English sentence from a movie review dataset \cite{Pang+Lee:04a} \\
        6 & as 5 but with rearranged words rendering them incorrect         \\
        \hline
    \end{tabular}
\end{center}
\centerline{ Table 1: Task Item Overview}
        
\subsection{Initial Results}

The following table shows the first analysis of task performance and response time for the items of the experiment task.\\
\begin{center}
    \begin{tabular}{ | l | l | }
        \hline
        Participants                         &   41                                           \\
        Gender                               &   20 Female / 21 Male                          \\
        Age                                  &   18-37 (Mean: 23.3, \textit{2 unreported})    \\
        \hline
        Average Task Performance (120 Items) &   111.7 (SD=5.72)  \\   
        Average Task Performance - Items 1-4 &   75.51 (SD=3.91)  \\   
        Average Task Performance - Items 5-6 &   36.19 (SD=2.54)  \\   
        Mean Task Response Time - Items 1-4  &   1.03s (SD=0.34)  \\
        Mean Task Response Time - Items 5-6  &   4.03s (SD=1.60)  \\
        \hline
    \end{tabular}
\end{center}
\centerline{Table 2: Initial results with participant's demographics}
\ \\

We are currently creating the dataset that will be comprised of eye and front camera video and the pupil dilation data from the eye tracker, the data from the heart rate monitor, and the video from a webcam facing the participants. Due to in-completion of the dataset, we are not reporting the results in this paper but give an overview of some of its properties.

\subsection{Conclusion and Future Work}

In the paper, we presented the initial results of our study conducted to understand the relationship between various physiological behaviours (HR, HRV, BR, PD) during an HCI task. We are currently analysing the data on all these behaviours, and our long-term goal lies in the development of a system that could potentially provide a robust non-intrusive measurement of mental workload in real-time during HCI. Our work is focused on creating a dataset of these physiological behaviours and later using them for a regression-based model to estimate mental load during various HCI and HRI scenarios.

\begin{acks}
We acknowledge funding and support from the EPSRC ORCA Hub (EP/R026173/1, 2017-2021) and consortium partners.
\end{acks}

\bibliographystyle{ACM-Reference-Format}
\bibliography{sample-sigchi-a}

\end{document}